\documentclass[12pt]{article}
\RequirePackage{cite}
\usepackage{times}
\usepackage{fullpage}
\usepackage{ifthen}
\usepackage{graphicx}
\usepackage{amsmath}
\usepackage{amssymb}
\usepackage[utf8]{inputenc}  
\usepackage{color}
\usepackage[usenames,dvipsnames]{xcolor}
\usepackage{lineno} 

\date{\today}

\makeatletter

\renewcommand\@biblabel[1]{#1.}

\def\@cite#1#2{$^{\mbox{\scriptsize #1\if@tempswa , #2\fi}}$}

\newcommand{\spacing}[1]{\renewcommand{\baselinestretch}{#1}\large\normalsize}
\spacing{2}

\def\@maketitle{%
  \newpage\spacing{1}\setlength{\parskip}{12pt}%
    {\Large\bfseries\noindent \textsf{\@title} \par}%
    {\noindent\sloppy \@author}%
}

\newenvironment{affiliations}{%
    \setcounter{enumi}{1}%
    \setlength{\parindent}{0in}%
    \slshape\sloppy%
    \begin{list}{\upshape$^{\arabic{enumi}}$}{%
        \usecounter{enumi}%
        \setlength{\leftmargin}{0in}%
        \setlength{\topsep}{0in}%
        \setlength{\labelsep}{0in}%
        \setlength{\labelwidth}{0in}%
        \setlength{\listparindent}{0in}%
        \setlength{\itemsep}{0ex}%
        \setlength{\parsep}{0in}%
        }
    }{\end{list}\par\vspace{12pt}}

\renewenvironment{abstract}{%
    \setlength{\parindent}{0in}%
    \setlength{\parskip}{0in}%
    \bfseries%
    }{\par\vspace{-6pt}}

\renewcommand{\section}{\@startsection {section}{1}{0pt}%
    {-6pt}{1pt}%
    {\bfseries}%
    }
\renewcommand{\subsection}{\@startsection {subsection}{2}{0pt}%
    {-0pt}{-0.5em}%
    {\bfseries}*%
    }

\newenvironment{methods}{%
    \section*{Methods}%
    \setlength{\parskip}{12pt}%
    }{}

\newenvironment{addendum}{%
    \setlength{\parindent}{0in}%
    \small%
    \begin{list}{Acknowledgements}{%
        \setlength{\leftmargin}{0in}%
        \setlength{\listparindent}{0in}%
        \setlength{\labelsep}{0em}%
        \setlength{\labelwidth}{0in}%
        \setlength{\itemsep}{12pt}%
        }
    }
    {\end{list}\normalsize}

\renewenvironment{figure}{\let\caption\NAT@figcaption}{}

\newcommand{\NAT@figcaption}[2][]{%
    \refstepcounter{figure}
    \par 
    \sffamily\noindent\textbf{Figure \arabic{figure}}\hspace{1em}#2
    }

\renewcommand{\subsection}{\@startsection {subsection}{2}{0pt}{-0pt}{-0.5em}{\textbf}*}

\makeatother

\title{Quantum-optical spectroscopy for plasma electric field measurements and diagnostics}

\author{David A. Anderson$^{1*}$, Georg Raithel$^{1, 2}$, Matthew Simons$^{3}$ and Christopher L. Holloway$^{3}$}

\usepackage[UKenglish]{babel}
\begin{document}

\maketitle

\begin{affiliations}
\item[] $^{1}$Rydberg Technologies, Ann Arbor, Michigan 48104, USA
\item[] $^{2}$Department of Physics, University of Michigan, Ann Arbor, Michigan 48109, USA
\item[] $^{3}$National Institute of Standards and Technology, Boulder, Colorado 80305, USA
\item[] $^*$Corresponding author.  E-mail: dave@rydbergtechnologies.com
\end{affiliations}

\begin{abstract}
Measurements of plasma electric fields are essential to the advancement of plasma science and applications.  Methods for non-invasive in situ measurements of plasma fields on sub-millimeter length scales with high sensitivity over a large field range remain an outstanding challenge.  Here, we introduce and demonstrate a new method for plasma electric field measurement that employs electromagnetically induced transparency as a high-resolution quantum-optical probe for the Stark energy level shifts of plasma-embedded Rydberg atoms, which serve as highly-sensitive field sensors with a large dynamic range.  The method is applied in diagnostics of plasmas photo-excited out of a cesium vapor.  The plasma electric fields are extracted from spatially-resolved measurements of field-induced shape changes and shifts of Rydberg resonances in rubidium tracer atoms.  Measurement capabilities over a range of plasma densities and temperatures are exploited to characterize plasmas in applied magnetic fields and to image electric-field distributions in cyclotron-heated plasmas.
\end{abstract}

The development of improved diagnostic techniques to measure the physical properties of plasmas is essential to advancing our understanding of fundamental plasma phenomena and plasma applications~\cite{Hutchinson.2002}.  Measurements of the electric and magnetic fields in plasmas are of particular importance to determining plasma parameters and investigating plasma processes in a broad range of plasma systems~\cite{Kimitaka.1996,Marklund.2001,Bale.2005,Rygg.2008,Li.2008}.  Existing optical diagnostic tools for direct measurements of plasma electric fields are largely based on emission spectroscopy and laser-induced fluorescence measurements of Stark shifts or other field-induced effects, such as level mixing in plasma-embedded atoms or molecules~\cite{Donnelly.1984,Moore.1984,Gavrilenko.2006, Czarnetzki.2009,Gigosos.2014} and field-sensitive Rydberg states~\cite{Ganguly.1987, Czarnetzki.1999, Feldbaum.2002, Takizawa.2004}.  In applications, process control in micro-fabrication has benefitted from optical diagnostics that provide in-situ monitoring in plasma etching, thin-film deposition, and discharge plasmas for surface cleaning~\cite{PlasmaDiagnostics.1989}.  Electrostatic (Langmuir) probes are another widely used tool for measurements of plasma parameters and electric potentials to deduce plasma electric fields.  In aerospace applications, as an example, electrostatic probes are used to determine the electric fields causing ion acceleration and electron heating in Hall thrusters~\cite{Staack.2004}.  A general limitation of electrostatic probes for plasma measurements is the inherent perturbation of the plasma potentials, the particle densities and energies caused by the insertion of the probe.  In practice, this requires complex models of the interaction between the plasma and probe, as it relates to the current-voltage readout of the probe from which information on plasma parameters is obtained~\cite{Stangeby.1995}.  Measurements of plasma electric fields at sub-millimeter length scales, with high sensitivity and range remain a significant challenge.

Advances in the ability to optically interrogate and exploit quantum systems at the single-atom level for measurements of physical quantities at higher precision continue to fuel the development and emergence of novel quantum measurement and sensing technologies.  Seminal developments include microwave and optical atomic clocks~\cite{Essen.1955,Heavner.2014,Ludlow.2015}, highly-sensitive, position-resolved atomic and diamond magnetometers~\cite{Savukov.2005,Patton.2012,Taylor.2008}, and inertial sensors based on atom interferometry~\cite{Cronin.2009,Battelier.2016}.  For measurements of electric fields, atoms in Rydberg states with a single electron excited to a high-lying orbit are exquisitely field-sensitive,~\cite{Gallagher,Frey.1993} and hold great promise for quantum electric-field metrology and sensing applications.  In combination with quantum-optical spectroscopy of the Rydberg atoms by electromagnetically induced transparency (EIT)~\cite{Mohapatra.2007}, atom-based measurement methods for static and time-varying electric~\cite{Barredo.2013,Sedlacek.2012,Holloway2.2014,Gordon.2014,Anderson.2016,Miller.2016, Anderson.2017} and magnetic fields up to 1~Tesla~\cite{Whiting.2016,Ma.2017,Zhang.2017} can be devised that are practical, calibration-free and non-invasive. In this work, we demonstrate for the first time the use of EIT as a quantum-optical probe of atomic Rydberg states of tracer atoms for high-precision absolute-standard measurements of plasma electric fields.

\begin{figure}
\centering
\includegraphics[width=1\linewidth]{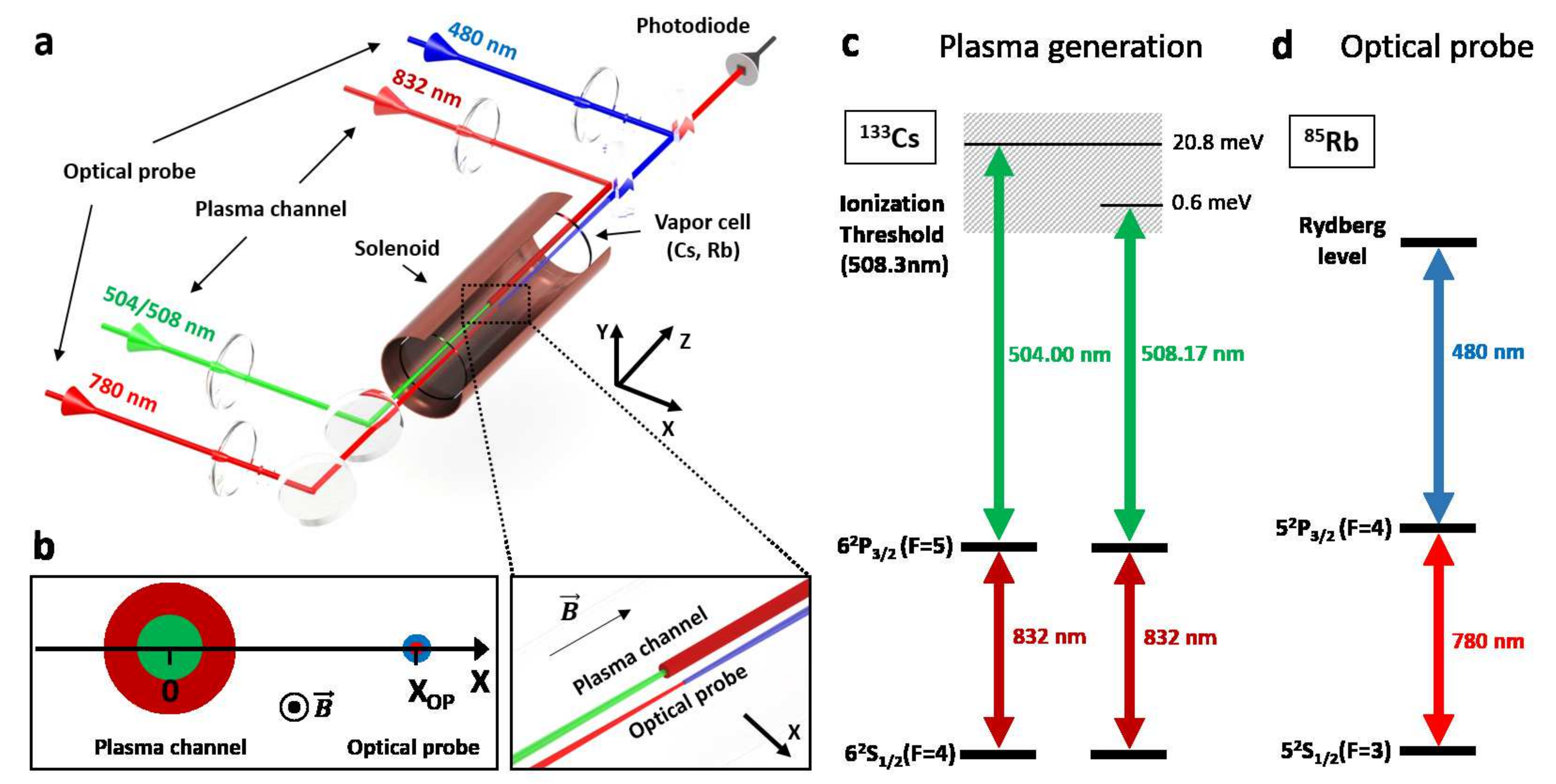}
\caption{\bf a \rm Experimental setup.  \bf b \rm Plasma channel and optical probe inside a dual species Cs and Rb vapor cell (right) and optical probe position $X_{\rm OP}$ relative to the plasma channel at $X=0$ (left); the magnetic-field direction is indicated. Single-atom energy-level diagrams for \bf c \rm Cs plasma photo-excitation with 0.6 meV and 20.8 meV electrons and \bf d \rm the Rb Rydberg-EIT optical probe.}
\label{fig:figure1}
\end{figure}

\noindent \bf Plasma generation and optical detection with Rydberg atoms. \rm The setup used for our measurements is illustrated in Fig.~\ref{fig:figure1}a.  We implement a dual-species spectroscopic cell containing both cesium (Cs) and rubidium (Rb) vapours.  Plasmas are generated by ionizing Cs atoms using continuous two-color photo-excitation with 852~nm and $\sim$510~nm laser beams counter-propagating and overlapped through the cell (Methods).  The optics for plasma generation are mounted on a dedicated platform for translation of the plasma channel along the $X$-axis across the cell relative to the optical probe beams fixed in position at $X_{OP}$, and propagating through the center of the cell (Fig.~\ref{fig:figure1}b). Figure~\ref{fig:figure1}c shows single-atom energy-level diagrams of Cs plasma photo-excitation.  The photo-ionization laser frequency in our experiments is set to within 0.01~nm to generate electron plasmas with initial kinetic energies of 0.6~meV (504~nm) or 20.8~meV (508.17~nm), allowing us to measure electric fields in two cases.

Plasma electric fields are measured by spectroscopically interrogating Rydberg states of $^{85}$Rb tracer atoms residing within the photo-excited Cs plasma using electromagnetically induced transparency (EIT) as an optical probe.  Single-atom energy-level diagrams of the Rydberg-EIT optical probe are shown in Fig.~\ref{fig:figure1}d.  Rydberg EIT is implemented using overlapped 780~nm and 480~nm laser beams, counter-propagating through the cell parallel to the plasma channel beams (Fig.~\ref{fig:figure1}a) (Methods). Relative cross sections of the beams in the optical probe and plasma channel are illustrated in Fig.~\ref{fig:figure1}b.

\begin{figure}
\centering
\includegraphics[width=1\linewidth]{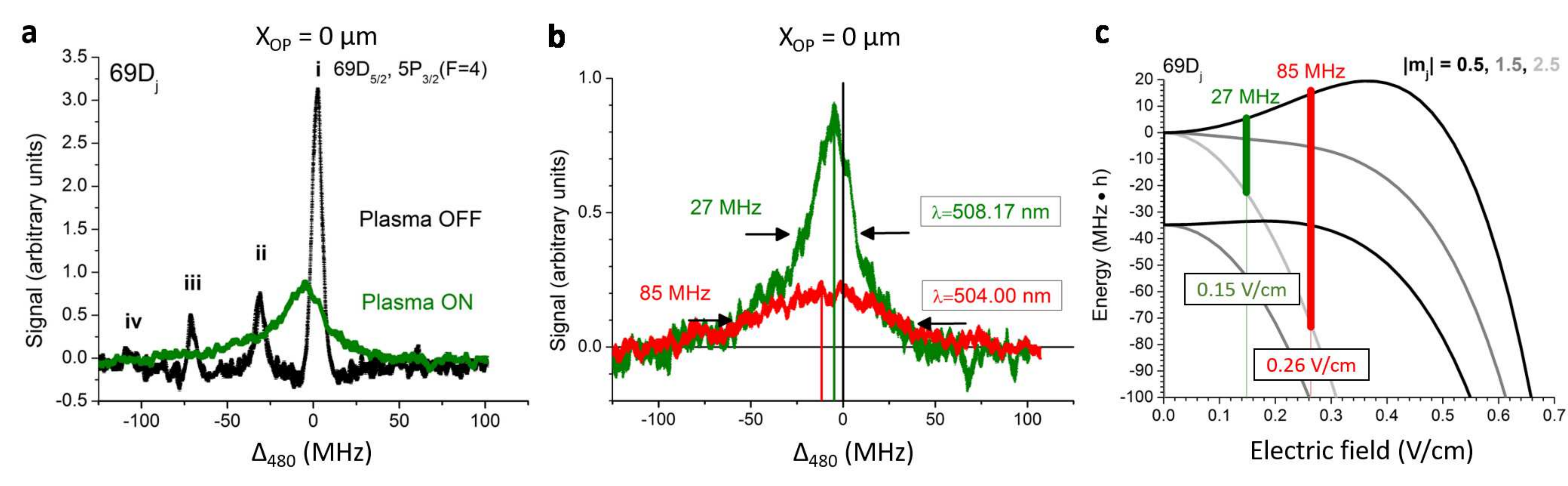}
\caption{Measured optical probe spectrum of the $^{85}Rb$ $69D_{J}$ Rydberg states for \bf a \rm without plasma (black) and with plasma (green) at $X_{\rm OP}$=0. The coupler detuning, $\Delta_{480}$, is relative to the field-free transition into the Rb $69D_{5/2}$ Rydberg state, used for electric-field sensing. The spectral lines i and ii correspond to the $J={5/2},{3/2}$ fine-structure states associated with the 780~nm transmission on the $5S_{1/2}(F=3)$ to $5P_{3/2}(F=4)$ transition; lines iii and iv correspond to states associated with the 780~nm transmission on the $5S_{1/2}(F=3)$ to $5P_{3/2}(F=3)$ transition. \bf b \rm Optical probe spectrum for plasma photo-excited with $\lambda=508.17$~nm (green) and $\lambda=$504~nm light (red).   \bf c \rm  Calculated Rydberg energy-level shift versus electric field.  The splitting between D$_{5/2}$ $\vert m_j\vert=$0.5 and 2.5 for measured half-widths in \bf b \rm and their electric fields (vertical lines).}
\label{fig:figure2}
\end{figure}

\noindent \bf Electric field measurements in photo-excited plasmas. \rm Plasma electric fields are derived from field-induced line-shape changes and level shifts of the plasma-embedded Rb atoms interrogated by the optical probe inside the cell.  An experimental $69D$ Rydberg spectrum without plasma (Cs photo-ionization beams off) is shown in black in Fig.~\ref{fig:figure2}a.  This spectrum is centered on the $69D_{5/2}$ Rydberg line of the Rb tracer atoms, observed in the form of an EIT window within the resonant $5S_{1/2}(F=3)$ to $5P_{3/2}(F=4)$ 780-nm transition of Rb.  The plasma-free EIT spectral line has a measured half-width of $2\pi\times 6.5$~MHz, limited by interaction-time broadening of the $^{85}$Rb atoms within the 780~nm probe beam.  

The spectra in green in Figs.~\ref{fig:figure2}a, b show the EIT readout from the optical probe along the plasma core ($X_{\rm OP}=$0) for a plasma with a low initial electron energy of 0.6~meV, obtained by photo-ionization of Cs with 508.17-nm green light.  The effect of the plasma electric fields is immediately evident.  Relative to the plasma-free case, the $69D_{5/2}$ line shifts by $-4.8\pm0.1$~MHz and broadens by more than a factor of four to a half-width of $2\pi\times 27$~MHz.  We determine the plasma field by comparing the broadening to calculated Rydberg Stark shifts for the Rb $69D$ $\vert m_j\vert$=0.5, 1.5, and 2.5 magnetic levels (Fig.~\ref{fig:figure2}c).  In the local frame of each tracer atom, the plasma electric field points in an arbitrary direction with respect to the optical probe laser polarizations, and therefore all $\vert m_j\vert$ contribute to the spectroscopic line.  The most probable plasma electric-field amplitude is determined by matching the half-width of the spectral line to the calculated splitting of the red- and blue-most shifted $\vert m_j\vert$-levels of the dominant $69D_{5/2}$ fine-structure component.  For the 0.6~meV plasma, we obtain an electric field $E=0.15$~V/m (green line in Fig.~\ref{fig:figure2}c).  Similarly, a spectrum for a plasma with an initial electron energy of 20.8~meV, obtained by photo-ionization of Cs at 504.00~nm, is shown in red in Fig.~\ref{fig:figure2}b.  There, the plasma induces a larger average shift of ($-11.4\pm0.1$)~MHz and broadening to $2\pi\times 85$~MHz, corresponding to an electric field $E=0.26$~V/cm (vertical red line in Fig.~\ref{fig:figure2}c). The field sensitivity reached in this measurement is an improvement of more than one order of magnitude compared to the highest sensitivity reached using other methods~\cite{Czarnetzki.1999, Feldbaum.2002, Takizawa.2004}.

The plasma electric fields are largely attributed to a steady-state positive space charge centered with the plasma axis.  Similar to a process observed in ultra-cold plasmas~\cite{Killian.1999}, electrons escape the plasma generation region, thereby building up a positive ionic space charge. The process continues until the space charge becomes large enough to slow the electron current and to promote a steady-state, outward-directed ion current that balances the electron current. In steady-state, the ion and electron currents are equal to the photo-ionization-induced charge production rate (assuming no recombination, which is a good approximation for our charge densities and temperatures, estimated below).  The $\sim 1.7 \times$-larger electric field measured in the 20.8~meV plasma is attributed to the greater initial kinetic electron energy, which translates into a deeper positive space-charge potential well necessary to slow the electron current so it can be balanced by the ion current.  Using Gauss's law, assuming a homogeneous charge density within the volume of the 780~nm beam, and taking its radius as the sensing radius, we estimate a net excess ionic charge density for the 20.8~meV plasma of $\rho=3\times10^7$~cm$^{-3}$.  From this we estimate a Debye length of $\lambda_D=150~\mu$m (electron and ion temperatures $T_e=T_i=300$~K, respectively) and plasma parameter $\Lambda =4\pi\rho\lambda_D^3 \geq 10^3$.  This is well within the plasma approximation ($\Lambda >> 1$) where collective effects dominate two-body processes, indicating the system may be considered a plasma.


The inhomogeneous plasma electric field has two components, the well-defined macroscopic field from the azimuthally symmetric plasma charge distribution, and the microscopic Holtsmark~\cite{Holtsmark.1919} electric field, which follows a distribution with a characteristic field strength of $2.603 n_i^{2/3} \vert e \vert / (4 \pi \epsilon_0)$, where $n_i$ is the ion density.  At any given location, the macroscopic field is mostly due to the combined effect of distant charges, while the Holtsmark field arises from the discreteness of the (randomly distributed) nearby charges. For the above ion density estimate, $\rho=3\times10^7$~cm$^{-3}$, the Holtsmark field $\sim 0.04$~V/cm. This is below our current level of field measurement precision (limited by the beam size), but appears accessible in future work.

\begin{figure}
\centering
\includegraphics[width=1\linewidth]{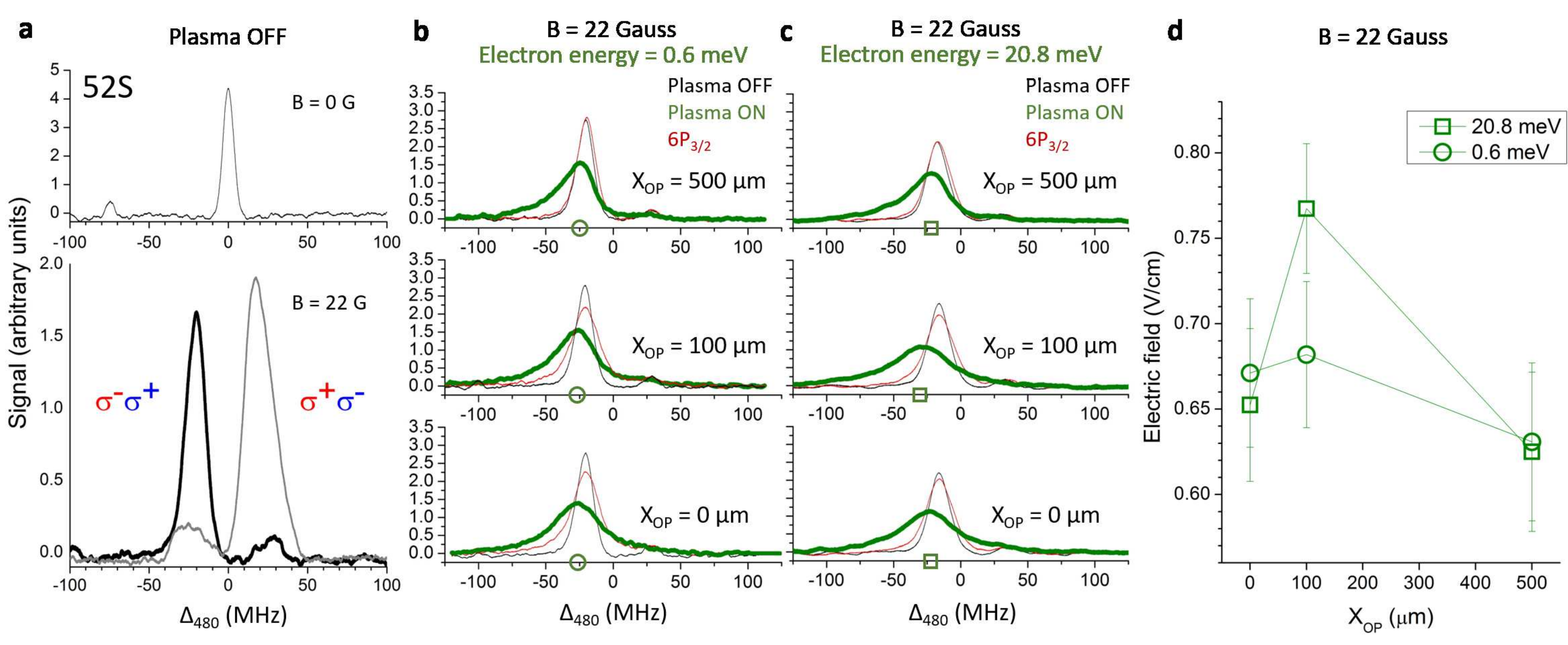}
\caption{\bf Electric-field measurements and spatial mapping in magnetically-confined plasmas. \rm  \bf a \rm Plasma-free 52S optical probe spectra without magnetic field (top) and with magnetic field (bottom) for indicated 780 nm (red) and 480 nm (blue) laser polarization configurations.  Optical probe spectra at $X_{\rm OP}=0, 100,$ and $500~\mu$m for 22~G magnetized plasmas with initial electron energies \bf b \rm 0.6~meV and \bf c \rm 20.8~meV.  Optical probe spectra are shown for the plasma off (black), plasma on (green), and photo-ionization light ($\lesssim 508$~nm) off (red).  \bf d \rm Plasma electric field corresponding to EIT peak shifts in \bf b \rm (circles) and \bf c \rm (squares); error bars reflect our $\pm 1$~MHz spectral resolution.}
\label{fig:figure3}
\end{figure}

\noindent \bf Electric field measurements of magnetically-confined plasmas. \rm  Rydberg EIT as a remote, non-invasive plasma diagnostic affords the ability to measure fields locally within the plasma on small spatial scales, at a resolution dictated by the EIT laser beam sizes, with a high electric-field measurement sensitivity and large dynamic range.  To demonstrate these capabilities and the utility of the method under a variety of plasma conditions, we measure and spatially map the electric fields of higher-density magnetically-confined plasmas.  In experiments, a homogeneous magnetic (B) field of tens of Gauss (1~Gauss$=10^{-4}$~Tesla) is applied along the plasma z-axis (Fig.~\ref{fig:figure1}a, Methods).  In the magnetic-field and kinetic-energy ($\lesssim k_B\times 300$~K) ranges of interest, the electron cyclotron radius is in the range of tens of micrometers while that of the ions is several meters.  We therefore expect considerable electron confinement, leading to larger plasma electric fields.  Here, we optically probe the plasma using the Rb 52S$_{1/2}$ Rydberg state.  This state offers a wider electric-field range that can be measured at the expense of lower sensitivity (DC polarizabilities of low-angular-momentum Rydberg states in alkali atoms ($\ell=0, 1, 2$ in Rb) scale as $\sim n^7$, and decrease at lower $\ell$-values~\cite{Gallagher}).  An S-state has the added advantage that it is without fine structure and has only two magnetic sub-levels ($m_j= -1/2, +1/2$), which exhibit the same Stark shifts. This simplifies the interpretation of the spectra in the magnetic field.


The Zeeman shifts of the Rb levels lead to polarization- and magnetic-field-dependent (plasma-free) optical probe spectra that are also affected by optical pumping~\cite{Zhang.2017}.  The effect is evident in Fig.~\ref{fig:figure3}a, which shows the plasma-free 52S$_{1/2}$ spectra with B=0~G (top panel) and 22~Gauss field for two configurations of circularly-polarized laser beams (bottom panel).  The optical pumping rates and resulting spectroscopic line shapes have been independently modelled using a quantum Monte-Carlo simulation and are well understood.  In our measurements, the $\sigma^-\sigma^+$ configuration is chosen because it results in a dominant, narrow EIT line that is red-shifted by the Zeeman effect. Since the electric-field-induced line shifts and broadenings are also to the red, the electric-field-induced features are spectrally isolated from the weak, blue-shifted Zeeman component in the EIT spectrum.

In Figs.~\ref{fig:figure3}b and c the optical probe is scanned to probe positions $X_{OP}$=0, 100, and 500~$\mu$m across plasmas photo-excited in a 22~G field with initial electron energies of 0.6~meV and 20.8~meV, respectively.  The plasma-free spectra and those for Cs($6P_{3/2}$) excitation with no photo-excitation laser are shown for reference.  The slight broadening observed for Cs(6P$_{3/2}$) excitation is attributed to a weak plasma resulting from photo-ionization of Cs(6P$_{3/2}$) atoms by the 480-nm Rb EIT readout beam. This EIT probe-induced ionization during the measurements is weak due to the lower intensity of the pulsed 480~nm light compared to that of the continuous $\lesssim 508$~nm light.  Also, the photo-ionization cross section of Cs(6P$_{3/2}$) at 480~nm is slightly lower than at $\lesssim 508$~nm.

From Figs.~\ref{fig:figure3}b, c and Fig.~\ref{fig:figure2}, we observe that the 52S line in the magnetized plasmas is substantially more broadened and red-shifted compared to the 69D line in the non-magnetized plasma.  Figure~\ref{fig:figure3}d gives radial electric-field profiles along X for the magnetized plasmas in Figs.~\ref{fig:figure3}b,c obtained using the measured peak shifts and 52S DC polarizability $\alpha=68.6$~MHz/(V/cm)$^2$. Along the core of the 0.6~meV magnetized plasma, the electric field is 0.67~V/cm, a factor of $\sim$4.5 higher than in the non-magnetized plasma in Fig.~\ref{fig:figure2}.  The larger electric fields in the magnetized plasma arise from electron confinement along the radial dimension. The electron cyclotron radius, $R_e=m_ev_e/eB \approx 35~\mu$m for a thermal electron velocity $v_e (T=300~K) \approx 13.5\times 10^3$~m/s, is smaller than other length scales of the system. Note that the Cs$^+$ cyclotron radius, $R_i=8.4$~m at 300~K, is so large that it is not important.  The electric field arises from the formation of a negative space-charge region centered along the photo-excitation beams by the pinning of the electrons to the magnetic field.  By symmetry, the electric field along the plasma core is small.  This is evident in the 20.8~meV-plasma field distribution (Fig.~\ref{fig:figure3}d), which has a field minimum at X$_{OP}=0~\mu$m.  Assuming that the negative space-charge distribution has an effective radial range on the order of the photo-ionization beam diameter, the electric field will increase as a function of radial distance $X_{OP}$ (for a homogeneous charge density the field would be $\sim X$).  Since outside of the space-charge distribution of the plasma the electric field is expected to drop as $X^{-1}$, the field should peak at a value $X$ on the order of the plasma radius. Figure~\ref{fig:figure3}d shows an electric-field maximum at $X_{OP}=100~\mu$m, which is slightly larger than the half-width half-maximum of the photo-ionization beam.


\begin{figure}
\centering
\includegraphics[width=1\linewidth]{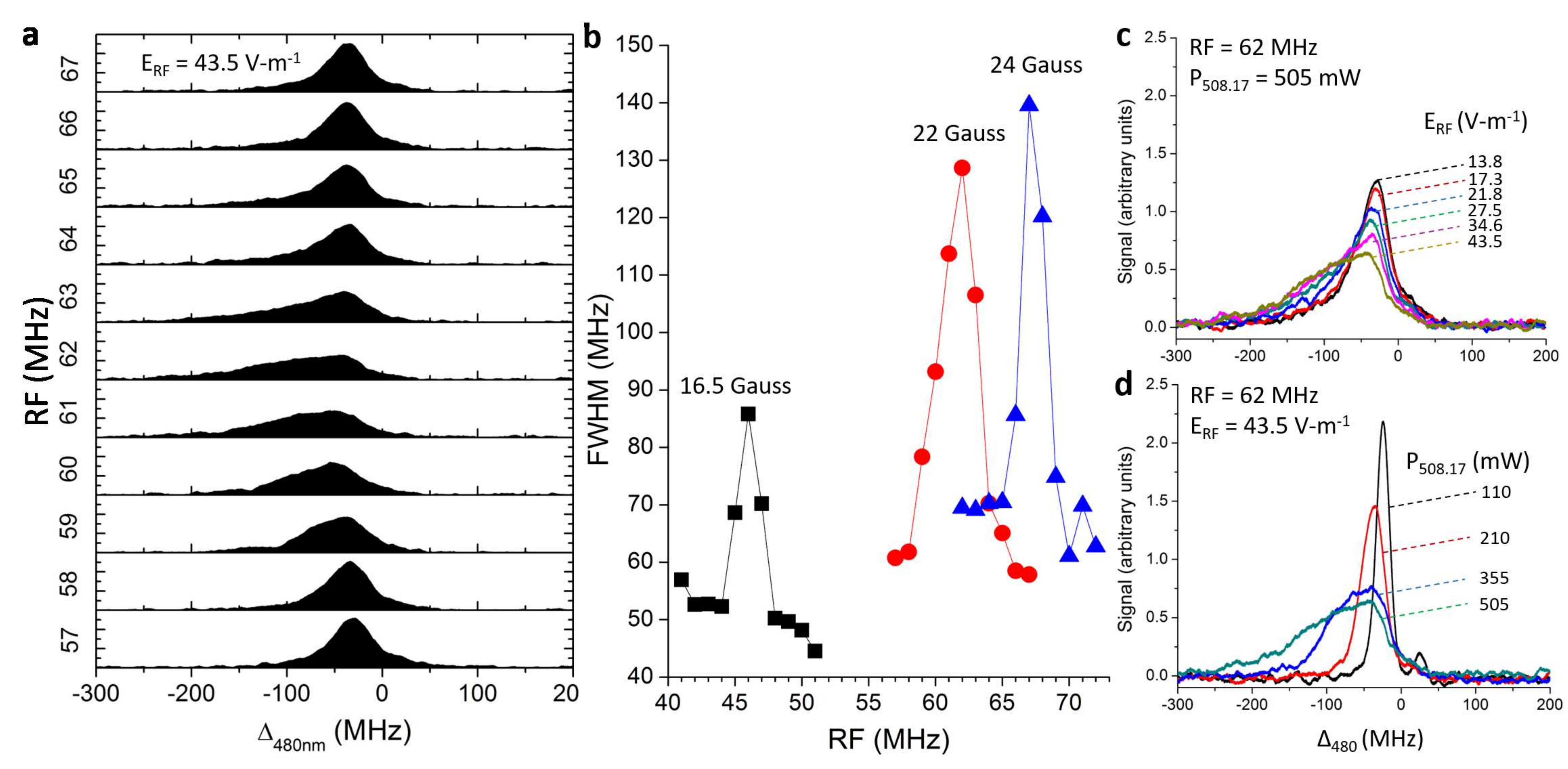}
\caption{\bf ECR heating of 508.17~nm photo-excited magnetized plasmas. \rm Optical probe spectra centered on the field-free $^{85}$Rb $52S_{1/2}$ Rydberg state at $X_{\rm OP}$=0 for \bf a \rm a series of RF field frequencies at fixed $E_{RF}$=43.5~V/m in a B=22~G magnetized plasma.  \bf b \rm ECRs for magnetized plasmas with B=16.5 (black squares), 22.0 (red circles), and 24.0~Gauss (blue triangles). Spectra for the B=22~G magnetized plasma resonantly driven with 62~MHz RF for \bf c \rm varying the RF field amplitude at a fixed power of 505~mW of the Cs photo-ionization laser (508.17-nm laser),  and \bf d \rm varying the 508.17~nm laser power at a fixed value of the RF electric field, $E_{RF}$=43.5~V/m.}
\label{fig:figure4}
\end{figure}

\noindent \bf \noindent \bf Electric field mapping of ECR plasmas. \rm  Resonance phenomena in plasmas provide fundamental insights into plasma processes and can in turn be exploited to control plasma properties by coupling the plasma to external fields that are resonant with a plasma mode.   In magnetized plasmas, electron cyclotron resonances (ECRs) are widely used for these purposes, ranging from basic studies of transport phenomena in magnetically-confined fusion plasmas~\cite{Bornatici.1983,Erckmann.1994} to heating of low-density and low-temperature plasmas in microfabrication~\cite{Asmussen.1989}.  To investigate resonant plasma behavior using EIT, we drive the ECRs of a 0.6~meV magnetized plasma with a resonant RF field $E_{RF}$ calibrated using RF-modulation spectroscopy~\cite{Miller.2016,Jiao.2016}. The field $E_{RF}$ is set such that the AC-Stark shift of the $52S$ Rydberg state is small compared to the shifts of interest due to the magnetic and plasma electric fields.

Figure~\ref{fig:figure4}a shows spectra of the 52S Rydberg line measured along the core of a 0.6~meV plasma magnetized by a 22~Gauss field for a series of equidistant RF field frequencies, centered on the ECR frequency $f_c=qB/2 \pi m_e$=62~MHz ($q$ is the elementary charge).  The line exhibits an increased asymmetric broadening and shift to lower energy as the RF field frequency approaches $f_c$=62~MHz.  The spectroscopic response to the ECR has a bandwidth of only several MHz and is well-centered at the expected ECR frequency.

To ensure the observed resonance is an ECR and not another unaccounted-for plasma resonance, the same measurement series is performed for several B-values.  Figure~\ref{fig:figure4}b shows the half-width of the optical-probe spectra versus RF field frequency for B=16.5, 22, and 24~G.  The observed resonances are peaked at 46, 62 and 67~MHz, respectively, in excellent agreement with the expected ECR frequency values.  We further verify the effect of the RF-drive field amplitude on the ECR resonance for B=22~Gauss (Fig.~\ref{fig:figure4}c).  As expected, the plasma electric field enhancement due to the ECR heating disappears as the drive field approaches zero.  We also vary the ionization laser power while resonantly driving the ECR, confirming that the plasma electric-field-induced spectroscopic shifts diminish at lower plasma charge density (Fig.~\ref{fig:figure4}d).

\begin{figure}
\centering
\includegraphics[width=1\linewidth]{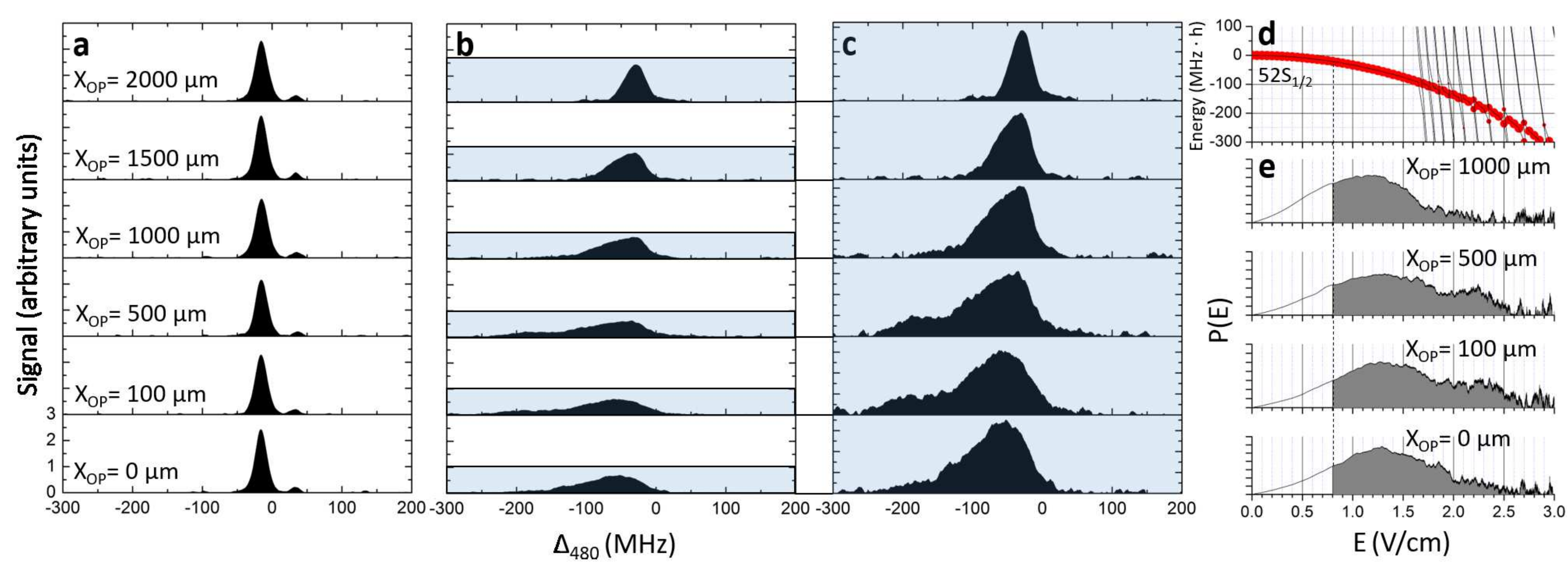}
\caption{Optical probe spectra at $X_{\rm OP}$ = 0, 100, 500, 1000, 1500, and 2000~$\mu$m with \bf a \rm no plasma at B=22~Gauss, \bf b \rm cyclotron-heated magnetized plasma at B=22~Gauss with photo-ionization wavelength 508.17~nm, RF field frequency 61.5~MHz, and RF field  $E_{RF}$=44.5~V/m, \bf c \rm normalized renderings of the spectra in \bf b \rm,  \bf d \rm calculated 52S$_{1/2}$ Stark map, and \bf e \rm electric field distributions $P(E)$ corresponding to the spectra in \bf b \rm and \bf c \rm for the indicated positions.}
\label{fig:figure5}
\end{figure}

It is of interest to measure electric-field distributions due to broadenings caused by micro-fields that may reveal the charged-particle density via the Holtsmark distribution~\cite{Holtsmark.1919}. For a set of radial positions $X_{\rm OP}$, spectra obtained in the 22~G ECR-heated plasma (Figs.~\ref{fig:figure5}b,c) are mapped over the known Stark shift of the 52S line (Fig.~\ref{fig:figure5}d) into an electric-field probability distribution $P(E)$ (Fig.~\ref{fig:figure5}e) (Methods). For $E\lesssim 0.8$~V/cm the values for $P(E)$ are considered inaccurate, because for small fields the Stark shift is less than the electric-field-free EIT line width (Fig.~\ref{fig:figure5}a).   The result shows considerable fields at $X_{\rm OP}=0$; these are interpreted as a combination of Holtsmark micro-fields and macroscopic fields in the wings of the probe beam (which extend up to $\sim 100~\mu$m away from the plasma symmetry axis, where macroscopic fields are non-zero). The distribution exhibits the largest fields of $~\sim$3~V/cm between $X_{\rm OP}=100$ and $500~\mu$m.  {The experiment in Figure~5 demonstrates that} the Rydberg-EIT probe can be used to acquire electric-field distributions $P(E)$.  Future applications will benefit from improved signal-to-noise. For instance, an improved signal-to-noise will allow for a more accurate deconvolution of the spectra that accounts for the EIT linewidth, and it may enable a study of the bimodal structure apparent in some $P(E)$ curves.

\noindent \bf Conclusion. \rm We have described a new measurement method of plasma electric fields that employs EIT as a quantum-optical probe for the Stark shifts of plasma-embedded Rydberg tracer atoms.  Plasma electric fields were measured down to 0.15~V/m, an improvement in sensitivity of more than an order of magnitude compared to traditional methods.  The applicability of the method to different plasma systems and plasma conditions was demonstrated in low-density plasmas, as well as in magnetized plasmas that have stronger electric fields due to electron trapping. We performed an initial study of plasma-field imaging, and we have analyzed ECR heating in our plasma system.  Our work paves the way for detailed investigations of plasma wave phenomena and transport processes.  The method could be employed in the determination of macroscopic plasma parameters in a wide range of systems, including higher density plasmas, where shorter-range interactions between tracer atoms and plasma constituents become relevant.  Electric fields above MV/m could in principle be measured, for example, by employing EIT with lower-lying Rydberg states or other EIT configurations.  With appropriate choices of tracer atoms, molecules or ions, the method may be extended to diagnostics in high-density and temperature plasmas relevant to inertial-confinement-fusion and magnetic-fusion plasma research and development.

\begin{methods}

\subsection{Experimental setup.}
The Cs plasmas are generated by two-stage photo-ionization using a 852~nm beam (420~$\mu$m full-width half-maximum of the intensity (FWHM) along the length of the cell, power of 2.0~mW), and a counter-propagating 510~nm laser beam (focused to 160~$\mu$m FWHM at the center of the cell, power 526~mW).  The 852~nm laser frequency is stabilized to the $62S_{1/2}(F=4)$ to $62P_{3/2}(F=5)$ transition; the 510~nm laser frequency is set to ionize out of the $6P_{3/2}(F=5)$ level (the photo-ionization threshold is at $508.3$~nm~\cite{Haq.2010}).

The quantum-optical plasma field probe employs EIT on Rb tracer atoms with a 480~nm beam, focused to a 85~$\mu$m FWHM and a power of 60~mW, and a 780~nm beam, focused to a 56~$\mu$m FWHM and has a power of 600~nW.  Rubidium Rydberg spectra are obtained by measuring the transmission of the 780~nm beam, whose frequency is stabilized to the $5S_{1/2}(F=3)$ to $5P_{3/2}(F=4)$ transition, while the 480~nm laser frequency is scanned linearly at a repetition rate of several Hz across a chosen Rydberg resonance.  All lasers have linewidths $<$1~MHz.

To magnetize the photo-excited plasmas, the vapor cell is placed inside a solenoid to generate homogeneous magnetic fields along the beam propagation direction ($Z$-axis).  The coil is wrapped in mu-metal to shield the cell from external magnetic fields.  The plasma is intrinsically shielded from external DC electric fields due to the dielectric vapor-cell walls~\cite{Mohapatra.2007,Miller.2016}.  Radio-frequency fields are generated by application of an RF signal to copper-film electrodes that are attached to the exterior of the cell.

\subsection{Electric field distributions}
The electric field distributions for the ECR-heated plasma in Fig.~\ref{fig:figure5}e are extracted from the optical probe spectra shown in Figs.~\ref{fig:figure5}b,c using the relation $P(E)=P(\nu)d\nu/dE$, where the distribution in frequency space is $P(\nu)$
and $d\nu/dE=\alpha_{52S}E$ ($\alpha_{52S}=68.6$~MHz/(V/cm)$^2$).  Before the computation of $P(E)$, a smoothing function was applied to the spectra $P(\nu)$, averaging over 20 steps in $\nu$. The smoothing corresponds to an electric-field averaging range of about 0.2~V/cm near E=0~V/cm to about 0.01 V/cm at E=3.0~V/cm.

\end{methods}

%



\begin{addendum}
 \item[Acknowledgements] This work was supported by Rydberg Technologies, the NSF (IIP-1624368 and PHY-1506093), and NIST through the Embedded Standards program.
 \item[Author Contributions] D.A.A. and G.R. conceived of the measurement method and experiments; D.A.A, G.R., M.S., and C.L.H. planned and performed the experiments; D.A.A. performed the data analysis.  D.A.A and G.R. wrote the paper.  All authors contributed to finalizing manuscript.
 \item[Author Information] Reprints and permissions information is available at www.nature.com/reprints. The authors declare no competing financial interests. Correspondence and requests for materials should be addressed to D.A.A.~(anderson.da@gmail.com).
\end{addendum}

\end{document}